\documentstyle[12pt]{article}

\def\Pom{{\bf I\!P}}

\def\lsim{\mathrel{\rlap{\lower4pt\hbox{\hskip1pt$\sim$}}
    \raise1pt\hbox{$<$}}}         

\def\gsim{\mathrel{\rlap{\lower4pt\hbox{\hskip1pt$\sim$}}
    \raise1pt\hbox{$>$}}}         

\begin{document}
\begin{center}
\phantom{.}
{\Large \bf Diffractive vector mesons \bigskip\\

beyond the s-channel helicity conservation}
\vspace{2.0cm}\\
{\large  E.V. Kuraev$^{1)}$, N.N. Nikolaev $^{2,3)}$
and B.G. Zakharov$^{3)}$,\vspace{0.5cm}\\}
{\sl
$^{1)}$Laboratory for Theoretical Physics, JINR, 141980 Dubna, Moscow
Region, Russia\medskip\\
$^{2}$IKP(Theorie), KFA J{\"u}lich, D-52428 J{\"u}lich, Germany \medskip\\
$^{3}$L. D. Landau Institute for Theoretical Physics, GSP-1,
117940, \\
ul. Kosygina 2, Moscow 117334, Russia}
\vspace{1.0cm}


{\bf Abstract}
\end{center}

We derive a full set, and determine the twist, of helicity amplitudes for 
diffractive production of light to heavy 
vector mesons in deep inelastic scattering. For 
large $Q^{2}$ all helicity amplitudes but the double-flip are calculable in 
perturbative QCD and are proportional to the gluon structure function 
of the proton at a similar hard scale. We find a substantial 
breaking of the $s$-channel helicity conservation which must persist 
also in real photoproduction.
\vspace{1.0cm}\\
Diffractive virtual photoproduction of vector mesons
$
\gamma^*+p\to V+p',
~~~V=\rho_0,\omega,\phi,J/\Psi,\Upsilon,
$
in deep inelastic scattering (DIS) at small $x={(Q^{2}+m_{V}^{2})}/
{(W^{2}+Q^{2})}$ is a testing ground of ideas on the QCD pomeron
exchange and light-cone wave function (LCWF) of vector mesons
(\cite{KNNZ93,NNZscan,Ginzburg,NNPZ96,NNPZZ98}, for the recent review
see \cite{Review}). (For the kinematics 
see fig.~1, $Q^{2}=-q^{2}$ and $W^{2}=(p+q)^{2}$ are standard
DIS variables). It is a self-analyzing process because the helicity 
amplitudes can be inferred from vector meson decay angular
distributions \cite{Wolf}. The property of $s$-channel helicity conservation 
(SCHC) and the dominance of transitions  $\gamma^*_L\to V_L$ and 
$\gamma^*_T\to V_T$ with $R=\sigma_L/\sigma_T \approx Q^2/m_V^2$  are 
shared by nearly all models ($L$ and $T$ stand for
the longitudinal and transverse polarizations). The nonperturbative
contributions to $\sigma_{T}$ tame the rise of $R$ with $Q^{2}$
\cite{NNZscan,NNPZ96}, but still the theoretical predictions seem to
exceed systematically the experimental evaluations of $R$. However,
these experimental data analyses suffer from an unwarranted assumption
of an exact SCHC \cite{Review}, which needs a theoretical scrutiny.

We report here a derivation  of the full set of helicity amplitudes
for transitions $\gamma^*_L\to V_L,\gamma^*_L\to V_T, \gamma^*_T\to V_T,
\gamma^*_T\to V_L$ for all flavours and 
small to moderate momentum transfer $\vec{\Delta}$
within the diffraction cone. We find substantial $s$-channel helicity 
non-conserving (SCHNC) effects similar to the SCHNC $LT$ interference 
found earlier by Pronyaev and two of the present authors (NNN and BGZ) 
for diffraction $\gamma^{*}p\rightarrow p'X$ into continuum states 
$X$ \cite{DIS97}.

The leading log${1\over x}$ (LL${1\over x}$) pQCD diagrams for vector
meson production are shown in Fig.~1. We treat vector mesons as $q\bar{q}$
states with the $V\bar{q}q$ vertex $\Gamma_{V} V_{\mu}\bar{u}\gamma_{\mu}u$.
At small $x$ it is sufficient to compute the imaginary part of the
amplitude, the real part is a small correction which can readily be
reconstructed from analyticity \cite{NNPZ96} and we do not discuss it
any more. We use the standard Sudakov expansion of all the momenta in 
the two lightcone vectors $$p'=p-q\frac{p^2}{s},\quad q'=q+p'\frac{Q^2}{s}$$
such that $q'^2= p'^2=0$ and $s=2p'\cdot q'$, and the two-dimensional
transverse component:
$k =z q'+y p'+k_\bot,\quad \kappa=\alpha q'+\beta p'+\kappa_\bot, \quad
\Delta=\gamma p'+\delta q'+ \Delta_\bot\,$ and for the final vector meson
$q_{V}=q+\Delta=q'+{m_{V}^{2}+\vec{\Delta}^{2} \over s}p'+\Delta_{\perp}$
(hereafter $\vec{k},\vec{\Delta},..$ always stand for $k_{\perp},
\Delta_{\perp}$ etc.). As usual, only the so-called
nonsense components in the Gribov's decomposition of gluon propagators
do contribute in the high-energy limit,
$$
D_{\mu\nu}(k)={2p'_{\mu} q'_{\nu} \over sk^2}\,,
$$
so that in the upper blob the amplitude $R_{\mu\nu\rho\sigma}$ of the
subprocess $g_{\mu}\gamma^{*}_{\sigma}\rightarrow q\bar{q} \rightarrow 
g'_{\nu}V_{\rho}$ enters in the form $I(\gamma^{*}\to V)\propto
p'_{\mu}p'_{\nu}R_{\mu\nu\rho\sigma}V^{*}_{\rho}e_{\sigma}$.
The vertex function $\Gamma$ is related to the radial LCWF of the $q\bar{q}$
Fock state of the vector meson as 
\begin{equation}
\psi_{V}(z,\vec{k}) =
{\Gamma_{V}(z,\vec{k}) \over
D(m_{V}^{2},z,\vec{k})}\,,
\label{eq:1}
\end{equation}
where $D(\vec{k}^2)=\vec{k}^2+m_{q}^2-z(1-z)m_{V}^2$.
For the parameterizations of LCWF's of vector mesons based on the technique
\cite{Berestezky} see \cite{NNPZ96}. The corresponding quantity for photons,
$
\psi_{\gamma}(z,\vec{k})= {1/D(-Q^{2},z,\vec{k})}\,$
only
differs by the substitution $\Gamma_{\gamma}(z,\vec{k})=1$ and $
m_{V}^{2} \to -Q^{2}$. To the LL${1\over x}$
the lower blob is related to the unintegrated gluon density matrix
${\cal{F}}(x,\vec{\kappa},\vec{\Delta})$ \cite{NNPZZ98,NZsplit,Lipatov}.

After standard elimination of the Sudakov parameters $y,\beta$ etc. from 
the on-mass shell condition, for instance, $(k-q)^{2}=-sy(1-z)-\vec{k}^{2}
-Q^2(1-z)=m_{q}^{2}$, the virtual photoproduction amplitude takes the form
\begin{eqnarray}
A(x,Q^{2},\vec{\Delta})=
\nonumber\\ 
is{C_{F} N_{c} c_{V}\sqrt{4\pi\alpha_{em}} 
\over 2\pi^{2}}
\int_{0}^{1} dz \int d^2\vec{k} \int {d^{2}\vec{\kappa}
\over
\kappa^{4}}\alpha_{S}({\rm max}\left\{\kappa^{2},
\vec{k}^{2}+m_{V}^{2}\right\}){\cal{F}}(x,\vec{\kappa},\vec{\Delta})
I(\gamma^{*}\to V)\, ,
\label{eq:2}
\end{eqnarray}
where $N_{c}=3$ is the number of colors, $C_{F}={N_{c}^{2}-1 \over 2N_{c} }$ 
is the Casimir operator, $C_{V}={1\over \sqrt{2}},
{1\over 3\sqrt{2}},{1\over 3},{2\over 3}$ for the 
$\rho^{0},\omega^{0},\phi^{0},J/\Psi$ mesons, $\alpha_{S}$ and $\alpha_{em}$
are the strong and electromagnetic couplings, respectively.
The dependence of ${\cal{F}}(x,\vec{\kappa},\vec{\Delta})$ on the variable 
$\vec{\Delta}\vec{\kappa}$ corresponds to the subleading BFKL singularities
\cite{Lipatov} and can be neglected.  For small $\vec{\Delta}$ within the 
diffraction cone
\begin{equation}
{\cal{F}}(x,\vec{\kappa},\vec{\Delta})=
{\partial G(x,\kappa^{2})\over \partial \log \kappa^{2}}
\exp(-{1\over 2}
B_{3\Pom}\vec{\Delta}^{2})\,.
\label{eq:3}
\end{equation}
where  $\partial G/\partial \log \kappa^{2}$ is the conventional
unintegrated gluon structure function and, modulo to a slow Regge growth,
the diffraction cone $B_{3\Pom}\sim$ 6 GeV$^{-2}$ \cite{NNPZZ98}.
The familiar polarization vectors for virtual photons are
$$
e_L=\frac{1}{Q}\left(q'+\frac{Q^2}{s}p'\right),~~
e_T=e_\bot.
$$
whereas for vector mesons
\begin{eqnarray}
V_{T}=V_\bot +{2 (\vec{V}_{\bot}\cdot \vec{\Delta}) \over s}(p'-q'),~~
V_{L}={1\over m_{V}}\left(q'+ {\vec{\Delta}^{2}-m_{V}^{2} \over s}p'
+\Delta_\bot\right)\, ,
\label{eq:4}
\end{eqnarray}
such that $(V_{T}V_{L})=(V_{T}q_{V})=(V_{L}q_{V})=0$.
Because of the small factor $1/s$ it is tempting to neglect
the component $\propto p'$ in $V_{T}$ but that would have been 
entirely erroneous. Indeed, closer inspection of the evaluation of
$I(\gamma^{*}\to V)$ shows that 
$p'_{\mu}p'_{\nu}R_{\mu\nu\rho\sigma}p'_{\rho}e_{\sigma}
\propto s p'_{\mu}p'_{\nu}R_{\mu\nu\rho\sigma}V^{*}_{\bot,\rho}e_{\sigma}$.
The same is true of the contribution from the component 
$\propto \Delta_\bot$ in $V_{L}$.

It is convenient to define $\vec{\Psi}_{\gamma,V}(\vec{k})
=\vec{k}\psi_{\gamma,V}(z,\vec{k})\,, \chi_{\gamma}(z,\vec{k})=
\left[\vec{k}^2+m_{q}^2- z(1-z)Q^2 \right]
\psi_{\gamma}(z,\vec{k})$,   $\chi_{V}(z,\vec{k})=
\left[\vec{k}^2+m_{q}^2+ z(1-z)m_{V}^2 \right]
\psi_{V}(z,\vec{k})$ and 
\begin{eqnarray}
\eta_{\gamma}=\psi_{\gamma}(z,\vec{k})-
\psi_{\gamma}(z,\vec{k}-\vec{\kappa}+\frac{1}{2}\vec{\Delta})
\, ,
\label{eq:5}\\
\eta_{V}=\psi_{V}(z,\vec{k}+(1-z)\vec{\Delta})-
\psi_{V}(z,\vec{k}-\vec{\kappa}+\frac{1}{2}(1-2z)\vec{\Delta})
\, ,
\label{eq:6}\\
\varphi_{\gamma}=
\chi_{\gamma}(z,\vec{k})-
\chi_{\gamma}(z,\vec{k}-\vec{\kappa}+\frac{1}{2}\vec{\Delta})
= -2z(1-z)Q^{2}\eta_{\gamma}\,
\label{eq:7}\\
\varphi_{V}=\chi_{V}(z,\vec{k}+(1-z)\vec{\Delta})-
\chi_{V}(z,\vec{k}-\vec{\kappa}+\frac{1}{2}(1-2z)\vec{\Delta})
\label{eq:8}\\
\vec{\phi}_{\gamma}=
\vec{\Psi}_{\gamma}(z,\vec{k})-
\vec{\Psi}_{\gamma}(z,\vec{k}-\vec{\kappa}+\frac{1}{2}\vec{\Delta}) 
\, ,
\label{eq:9}\\
\vec{\phi}_{V}=
\vec{\Psi}_{V}(z,\vec{k}+(1-z)\vec{\Delta})-
\vec{\Psi}_{V}(z,\vec{k}-\vec{\kappa}+\frac{1}{2}(1-2z)\vec{\Delta})
\, ,
\label{eq:10}\\
\Phi_{2}=\psi_{\gamma}(z,\vec{k}-(1-z)\vec{\Delta})-
\psi_{\gamma}(z,\vec{k}-\vec{\kappa}-\frac{1}{2}(1-2z)\vec{\Delta})
\nonumber\\
-\psi_{\gamma}(z,\vec{k}+\vec{\kappa}-\frac{1}{2}(1-2z)\vec{\Delta})+
\psi_{\gamma}(z,\vec{k}+z\vec{\Delta})\, ,
\label{eq:11}\\
\vec{\Phi}_{1}=\vec{\Psi}_{\gamma}(z,\vec{k}-(1-z)\vec{\Delta})-
\vec{\Psi}_{\gamma}(z,\vec{k}-\vec{\kappa}-\frac{1}{2}(1-2z)\vec{\Delta})
\nonumber\\
-
\vec{\Psi}_{\gamma}(z,\vec{k}+\vec{\kappa}-\frac{1}{2}(1-2z)\vec{\Delta})+
\vec{\Psi}_{\gamma}(z,\vec{k}+z\vec{\Delta})\, .
\label{eq:12}
\end{eqnarray}
Then the integrands $I(\gamma^{*}\to V)$ in (\ref{eq:2}) can be cast
in the form which emphasizes the $V\leftrightarrow \gamma$ symmetry nicely:
\begin{eqnarray}
I(\gamma^{*}_L \to V_{L})={1\over Q m_{V}}\varphi_{\gamma}\varphi_{V}=
-{2Q \over m_{V}}z(1-z)\Phi_{2}
\left[m_{q}^{2}+\vec{k}^{2}+z(1-z)m_{V}^{2}\right]
\psi_{V}(z,\vec{k})
\, ,
\label{eq:13}\\
I(\gamma^{*}_T \to V_{T}) =(\vec{V}^{*}\vec{e})[m_{q}^{2}
\eta_{V}\eta_{\gamma}+(\vec{\phi}_{V}\vec{\phi}_{\gamma})]
+(1-2z)^{2}(\vec{\phi}_{\gamma}\vec{e})(\vec{\phi}_{V}\vec{V}^{*})
-(\vec{\phi}_{V}\vec{e})(\vec{\phi}_{\gamma}\vec{V}^{*})\nonumber\\
 =\left\{(\vec{V}^{*}\vec{e})[m_{q}^{2}\Phi_{2}+
(\vec{k}\vec{\Phi}_{1})]
+(1-2z)^{2}(\vec{V}^{*}\vec{k})(\vec{e}\vec{\Phi}_{1})
-(\vec{e}\vec{k})(\vec{V}^{*}\vec{\Phi}_{1})\right\}\psi_{V}(z,\vec{k})
\, ,
\label{eq:14}\\
I(\gamma^{*}_T \to V_{L})=\frac{(1-2z)}{m_{V}}(\vec{\phi}_{\gamma}
\vec{e})\varphi_{V}= \frac{(1-2z)}{m_{V}}
(\vec{e}\vec{\Phi}_{1})\left[m_{q}^{2}+\vec{k}^{2}+z(1-z)m_{V}^{2}\right]
\psi_{V}(z,\vec{k})\, ,
\label{eq:15} \\
I(\gamma^{*}_L \to V_{T})={(1-2z)\over Q}}\varphi_{\gamma}
({\vec{V^*}\vec{\phi}_{V})= -2Qz(1-z)(1-2z)(\vec{V}^{*}\vec{k})\Phi_{2}
\psi_{V}(z,\vec{k})\,.
\label{eq:16}
\end{eqnarray}
We consistently keep the quark mass so that our derivation holds from light
to heavy vector mesons.

After simple shifts of the integration variables to the common argument 
of the vector meson LCWF, the $I(\gamma^{*}\to V)$
can also be cast in the second form shown in (\ref{eq:13})-(\ref{eq:16}).
This second form emphasizes the close analogy between amplitudes of
vector meson production and diffraction into continuum discussed in
\cite{DIS97,NZsplit,NPZslope}, the major difference being the emergence of
$\psi_{V}(z,\vec{k})$ instead of the plane WF. The second
form is also convenient for the derivation of the $1/Q^{2}$ expansion,
i.e., of the twist, of helicity 
amplitudes. To this end we notice that $\psi_{V}(z,\vec{k})$ varies
on a hadronic scale $k^{2} \sim R_{V}^{-2}$, where $R_{V}$ is a radius
of the vector meson, whereas $\psi_{\gamma}(z,\vec{k})$ is a slow
function which varies on a large pQCD scale
\begin{equation}
\overline{Q}^{2}=m_{q}^{2} + z(1-z)Q^{2}\, ,
\label{eq:17}
\end{equation}
which allows a systematic twist expansion in powers of $1/\overline{Q}^{2}$
for $\Delta$ within the diffraction cone. In all cases but the double
helicity flip the dominant twist amplitudes come from the leading 
log$\overline{Q}^{2}$ (LL$\overline{Q}^{2}$) region of $\vec{k}^{2}\sim 
R_{V}^{-2},\vec{\Delta}^{2} \ll \vec{\kappa}^{2} \ll \overline{Q}^{2}$. 
After some algebra in (\ref{eq:13})-(\ref{eq:16}) we find
\begin{eqnarray}
I(\gamma^{*}_{L} \to V_{L})=-{Q \over m_{V}}\cdot
{4z(1-z)[m_q^2+\vec{k}^2+z(1-z)m_V^2]\over \overline{Q}^4}
\cdot \psi_{V}(z,\vec{k})\vec{\kappa}^2\, ,
\label{eq:18}\\
I(\gamma^{*}_T \to V_{T};\lambda_{V}=\lambda_{\gamma})=
2(\vec{V}^*\vec{e})\cdot
{m_q^2+2[z^2+(1-z)^2]\vec{k}^2 \over\overline{Q}^4}\cdot\psi_{V}(z,\vec{k})
\vec{\kappa}^2\, , 
\label{eq:19}\\
I(\gamma^{*}_L \to V_{T})=
-8{(\vec{V}^*\vec{\Delta})Q\over \overline{Q}^{2}}\cdot
{z(1-z)(1-2z)^2
\over \overline{Q}^4}\cdot\vec{k}^{2}\psi_{V}(z,\vec{k})\vec{\kappa}^2\, ,
 \label{eq:20}\\
I(\gamma^{*}_T \to V_{L})=-2{(\vec{e}\vec{\Delta})\over m_{V}}\cdot
{[m_q^2+\vec{k}^2+z(1-z)m_V^2]
(1-2z)^2
\over \overline{Q}^4}\cdot \psi_{V}(z,\vec{k})\vec{\kappa}^2\, .
 \label{eq:21}
\end{eqnarray}
Here we split explicitly the amplitude (\ref{eq:14}) into the
SCHC helicity-non-flip component (\ref{eq:19}) and SCHNC double-helicity 
flip component
\begin{eqnarray}
I(\gamma^{*}_T \to V_{T'};\lambda_{V}=-\lambda_{\gamma})=
2(\vec{V}^*\vec{\Delta})(\vec{e}\vec{\Delta})\cdot
{z(1-z) \over \overline{Q}^{4}}
\left[6(1-2z)^2{\vec{\kappa}^{2}\over\overline{Q}^{2}}+1\right]
\vec{k}^{2}\psi_{V}(z,\vec{k})\, .
 \label{eq:22}
\end{eqnarray}

In all cases when $I(\gamma^{*}\to V) \propto \kappa^{2}$ the  
LL$\overline{Q}^{2}$ approximation is at work, the gluon structure function 
enters the integrand in the form 
\begin{equation}
\int^{\overline{Q}^{2}} {d\kappa^{2}\over \kappa^{2}}
{\partial G(x,\kappa^{2})\over \partial \log \kappa^{2}} =
G(x,\overline{Q}^{2})
\label{eq:23}
\end{equation}
and after the $z$ integration one finds that the helicity amplitudes 
will be proportional to $\alpha_{S}(Q_{V}^{2})G(x,Q_{V}^{2})$,
where the pQCD hard scale 
\begin{equation}
Q_{V}^{2} \sim 
(0.1-0.2)(Q^{2}+m_{V}^{2})\, 
\label{eq:24}
\end{equation}
can depend on helicities because of the different end-point 
contributions from $z\rightarrow 0$ and $z\rightarrow 1$ where the 
running hardness $\overline{Q}^{2}$ is small \cite{NNZscan,NNPZ96}. 
This issue and the 
sensitivity of helicity amplitudes to the LCWF of vector mesons will 
be discussed elsewhere. Here we cite our results for the dominant twist 
SCHC and single-flip SCHNC amplitudes $A_{\lambda_V\lambda_{\gamma}}$ 
in the helicity basis:
\begin{eqnarray}
A_{0L} \propto {Q\over m_{V}}\cdot 
{\alpha_{S}(Q_{V}^{2})\over (Q^{2}+m_{V}^{2})^{2}}
\cdot G(x,Q_{V}^{2})\, ,
\label{eq:25}\\
A_{\pm\pm} \propto  {\alpha_{S}(Q_{V}^{2})\over (Q^{2}+m_{V}^{2})^{2}}
\cdot G(x,Q_{V}^{2}) \, ,
\label{eq:26}\\
A_{\pm L} \propto {\Delta \over m_{V}}
\cdot {\alpha_{S}(Q_{V}^{2})\over (Q^{2}+m_{V}^{2})^{2}}
\cdot G(x,Q_{V}^{2})\, ,
\label{eq:27}\\
A_{0 \pm} \propto {Q\over m_{V}}
\cdot{m_{V}\Delta\over Q^{2}+m_{V}^{2}}\cdot
{\alpha_{S}(Q_{V}^{2})\over (Q^{2}+m_{V}^{2})^{2}}\cdot G(x,Q_{V}^{2})\, ,
\label{eq:28}
\end{eqnarray}
First, all above amplitudes are pQCD calculable for large $Q^{2}$ and/or
heavy vector mesons. Second, the factor $Q/m_{V}$ in the longitudinal 
photon amplitudes (\ref{eq:25}) and (\ref{eq:27}) is a generic consequence 
of gauge invariance irrespective of the detailed production dynamics.  
Third, the longitudinal Fermi momentum of quarks is $k_{z}\sim {1\over 2}
m_{V}(2z-1)$ and eqs. (\ref{eq:20}),(\ref{eq:21}) make it obvious that 
single-helicity flip requires a longitudinal Fermi motion of 
quarks and vanishes in the nonrelativistic limit. Similarly, double-helicity
flip requires the transverse Fermi motion of quarks. Fourth, the leading  
SCHNC effect is an interference of $A_{0L}$ and $A_{0\pm}$  and the first
experimental indications for that have been reported recently 
\cite{Vancouver}. We hotice here that a duality consistency \cite{GNZlong} 
holds between the twist of $\propto  A_{0L}A_{0\pm}$ and that of the LT 
interference structure function $F^{D}_{LT}$ derived in \cite{DIS97}. 
The issue of duality for SCHNC diffractive DIS will be discussed elsewhere. 
Fifth, excitation of transverse mesons by longitudinal photons is 
of higher twist compared to excitation of longitudinal mesons 
by transverse photons.

In contrast to the above, the dominant twist double-helicity flip amplitude 
comes from the non-leading-log$\overline{Q}^{2}$ term 1 in the square 
brackets in (\ref{eq:22}) and is proportional to 
\begin{equation}
\int^{\overline{Q}^{2}} {d\kappa^{2}\over \kappa^{4}}
{\partial G(x,\kappa^{2})\over \partial \log \kappa^{2}} \sim
{1\over \mu_{G}^{2}} G(x,\mu_{G}^{2})\,
\label{eq:29}
\end{equation}
where a soft scale $\mu_{G} \sim $ 0.7-1 GeV is set by the inverse radius 
of propagation of perturbative gluons. Precisely the same nonperturbative 
quantity (\ref{eq:29}) describes the contribution from the $\gamma^{*}\to 
V$ transition vertex to the diffraction slope for helicity-non-flip 
amplitudes, for more discussion see \cite{NNPZZ98}. Then from 
eq.~(\ref{eq:21}) we find 
\begin{equation}
A_{\pm\mp} \propto \Delta ^{2}
\cdot {\alpha_{S}(Q_{V}^{2})\over (Q^{2}+m_{V}^{2})^{2}}
\left[ {6G(x,Q_{V}^{2})\over Q^{2}+m_{V}^{2}}\cdot{\langle k_{z}^{2} \rangle
\over 4m_{V}^{2}} 
+{G(x,\mu_{G}^{2})\over \mu_{G}^{2}} \right]\, ,
\label{eq:30}
\end{equation}
where the leading log$\overline{Q}^{2}$ amplitude is of higher twist. 
Such a mismatch of the twist and leading log$\overline{Q}^{2}$ regime
in diffractive DIS is d\'ej\`a vu: the leading twist $\sigma_{T}$ 
is soft-gluon 
dominated whereas the full fledged log$\overline{Q}^{2}$ is at work 
for the higher twist $\sigma_{L}$ \cite{NZsplit,GNZlong}. What is new in 
(\ref{eq:30}) is that the both regimes mix in one and the same helicity 
amplitude. The soft-gluon exchange dominance of the leading twist  
double-helicity flip was noticed recently by Ivanov and Kirschner 
\cite{Ivanov}. 

Finally, we emphasize that non-vanishing single- and double-helicity flip 
amplitudes (\ref{eq:27}) and (\ref{eq:29}) do not require the applicability 
of pQCD and can best be searched for in real photo- or electroproduction 
at small $Q^{2}\lsim m_{V}^{2}$.

When this manuscript was under preparation, we learned of the related
work by Ivanov and Kirschner (IK) \cite{Ivanov}. IK considered only
light quarkonia and put $m_{q}=0$, our results are applicable from light
to heavy quarkonia. While we agree with IK on the twist of helicity 
amplitudes, he have differences in the form if $I(\gamma^{*}\to V)$. 
For instance, the $\gamma^{*}\leftrightarrow V$ symmetry is not manifest 
in the IK formulas. The apparent source of differences is the unwarranted
omission by IK of terms $\propto \Delta$ in the polarization vectors of vector 
mesons.

To summarize, we presented the perturbative QCD derivation of 
helicity amplitudes for diffractive electroproduction of vector mesons.
Compared to IK \cite{Ivanov} our derivation holds for light to heavy
vector mesons and our formulas are applicable also to production of
radially excited states.  
We determined the twist of the $s$-channel helicity non-conserving 
amplitudes. With the exception of double-flip, all helicity amplitudes 
are proportional to the gluon structure function of the proton at a 
similar pQCD hardness scale (\ref{eq:24}). Our principal conclusions 
on substantial $s$-channel helicity non-conserving effects hold 
beyond the perturbative QCD and are applicable also to real photoproduction.
 
The authors are grateful to S.Gevorkyan for discussions, D.Ivanov for
useful correspondence on ref. \cite{Ivanov}, K.Piotrzkowski on
the information on presentations \cite{Vancouver} at the recent ICHEP-98 
in Vancouver and to I.Akushevich for pointing out a misprint. EAK 
is grateful to Institute f\"ur Kernphysik of 
Forschungszentrum J\"ulich
for the hospitality. The work of EAK and BGZ has been supported partly
by the INTAS Grants 93-0239 and 96-0597, respectively.

{\bf Figure caption:}\\

Fig.1: One of the four Feynman diagrams for the vector meson production
$\gamma^{*}p\rightarrow V p'$ via QCD two-gluon pomeron
exchange.

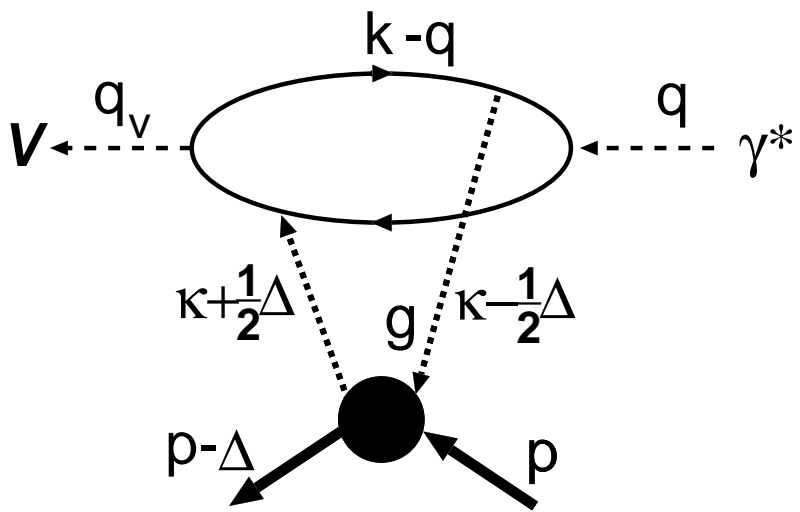
\end{document}